\documentclass[12pt]{article}

\usepackage{graphicx}
\usepackage{floatflt, graphicx}
\usepackage{amssymb}
\usepackage{subfigure}

\begin{document}

\title {\bf Cosmic Neutrinos and the Energy Budget of Galactic and Extragalactic Cosmic Rays%
\footnote{Talk presented at the International Workshop on Energy Budget in the High Energy Universe, Kashiwa, Japan, February 2006}%
}
\author{Francis Halzen\\
Department of Physics, University of Wisconsin, Madison, WI 53706}

\date{}

\maketitle

\vspace*{-.4in}

\begin{abstract}
Although kilometer-scale neutrino detectors such as IceCube are discovery instruments, their conceptual design is very much anchored to the observational fact that Nature produces protons and photons with energies in excess of $10^{20}$\,eV and $10^{13}$\,eV, respectively. The puzzle of where and how Nature accelerates the highest energy cosmic particles is unresolved almost a century after their discovery. We will discuss how the cosmic ray connection sets the scale of the anticipated cosmic neutrino fluxes. In this context, we discuss the first results of the completed AMANDA detector and the science reach of its extension, IceCube.\end{abstract}

\section{Introduction}

Ambitious projects have been launched to extend conventional astronomy beyond wavelengths of $10^{-14}$\,cm, or GeV photon energy. Besides gamma rays, protons (nuclei), neutrinos and gravitational waves will be explored as astronomical messengers probing the extreme Universe. The challenges are considerable:
\begin{itemize}
\item Protons are relatively abundant, but their arrival directions have been scrambled by magnetic fields.
\item $\gamma$-rays do point back to their sources, but are absorbed at TeV-energy and above on cosmic background radiation.
\item neutrinos propagate unabsorbed and without deflection throughout the Universe but are difficult to detect.
\end{itemize}
Therefore, multi-messenger astronomy may not just be an advantage, it may be a necessity for solving some of the outstanding problems of astronomy at the highest energies such as the identification of the sources of the cosmic rays, the mechanism(s) triggering gamma ray bursts and the particle nature of the dark matter.

We here discuss the case for the detection of neutrinos associated with the observed fluxes of high energy cosmic rays and gamma rays; it points, unfortunately, at the necessity of commissioning kilometer-scale neutrino detectors. Though ambitious, the scientific case is compelling because neutrinos will reveal the location of the source(s) and represent the ideal tool to study the black holes powering the cosmic accelerator(s).

Soon after the discovery in the mid-fifties that neutrinos were real particles and not just mathematical constructs of theorists' imagination, the idea emerged that they represent ideal cosmic messengers\cite{reines}. Because of their weak interactions, neutrinos reach us unimpeded from the edge of the Universe and from the inner reaches of black holes. The neutrino telescopes now under construction have the capability to detect neutrinos with energies from a threshold of $\sim 10$\,GeV to, possibly, $ \sim 10^2$\,EeV, the highest energies observed. Their telescope range spans more than 10 orders of magnitude in wavelengths smaller than $10^{-14}$\,cm. This is a reach equivalent to that of a hypothetical astronomical telescope sensitive to wavelengths from radio to X-rays. Above $10^5$\,TeV the observations are free of muon and neutrino backgrounds produced in cosmic ray interactions with the Earth's atmosphere. Each neutrino is a discovery.\footnote{We will use  GeV$=10^9$\,eV, TeV$=10^{12}$\,eV,  PeV$=10^{15}$\,eV and EeV$=10^{18}$\,eV units of energy}

The real challenge of neutrino astronomy is that kilometer-scale neutrino detectors are required to do the science. The first hint of the scale of neutrino telescopes emerged in the nineteen seventies from theoretical studies of the flux of neutrinos produced in the interactions of cosmic rays with microwave photons, the so-called Greissen-Zatsepin-Kuzmin or GZK neutrinos. Since then the case for kilometer-size instruments has been strengthened\cite{PR} and the possibility of commissioning such instruments demonstrated\cite{ice3}. In fact, if the neutrino sky were within reach of smaller instruments, it would by now have been revealed by the first-generation AMANDA telescope. It has been taking data since 2000 with a detector of $0.01 \sim 0.08$\,km$^2$ telescope area, depending on the sources\cite{pune}.

Given the size of the detector required, all efforts have concentrated on transforming large volumes of natural water or  ice into Cherenkov detectors. They reveal the secondary muons and electromagnetic and hadronic showers initiated in neutrino interactions inside or near the detector. Because of the long range of the muon, from kilometers in the TeV range to tens of kilometers at the highest energies, neutrino interactions can be identified far outside the instrumented volume. Adding to the technological challenge is the requirement that the detector be shielded from the abundant flux of cosmic ray muons by deployment at a depth of typically several kilometers. After the cancellation of a pioneering attempt\cite{water} to build a neutrino telescope off the coast of Hawaii, successful operation of a smaller instrument in Lake Baikal\cite{baikal} bodes well for several efforts to commission neutrino telescopes in the Mediterranean\cite{water,emigneco}. We will here mostly concentrate on the construction and first four years of operation of the AMANDA telescope\cite{pune,nature} which has transformed a large volume of natural deep Antarctic ice into a Cherenkov detector. It represents a first-generation telescope as envisaged by the DUMAND collaboration over 20 years ago and a proof of concept for the kilometer-scale IceCube detector, now under construction.

Even though neutrino ``telescopes" are designed as discovery instruments covering a large dynamic range, be it for particle physics or astrophysics, their conceptual design is very much anchored to the observational fact that Nature produces protons and photons with energies in excess of $10^{20}$\.eV and $10^{13}$\,eV, respectively. The cosmic ray connection sets the scale of cosmic neutrino fluxes. We will discuss this first.

\section{Cosmic Neutrinos Associated with Extragalactic Cosmic Rays}
 
Cosmic accelerators produce particles with energies in excess of $10^8$\,TeV; we do not know where or how. The flux of cosmic rays observed at Earth is sketched in Fig.\,1a,b\cite{gaisseramsterdam}. The  energy spectrum follows a broken power law. The two power laws are separated by a feature dubbed the ``knee"; see Fig.\,1a. Circumstantial evidence exists that cosmic rays, up to perhaps EeV energy, originate in galactic supernova remnants. Any association with our Galaxy disappears in the vicinity of a second feature in the spectrum referred to as the ``ankle". Above the ankle, the gyroradius of a proton in the galactic magnetic field exceeds the size of the Galaxy and it is generally assumed that we are  witnessing the onset of an extragalactic component in the spectrum that extends to energies beyond 100\,EeV. Experiments indicate that the highest energy cosmic rays are predominantly protons or, possibly, nuclei. Above a threshold of 50 EeV these protons interact with cosmic microwave photons and lose energy to pions before reaching our detectors. This is the GZK cutoff that limits the sources to our local supercluster. 

\begin{figure}[h]
\centering\leavevmode
\includegraphics[width=5.8in]{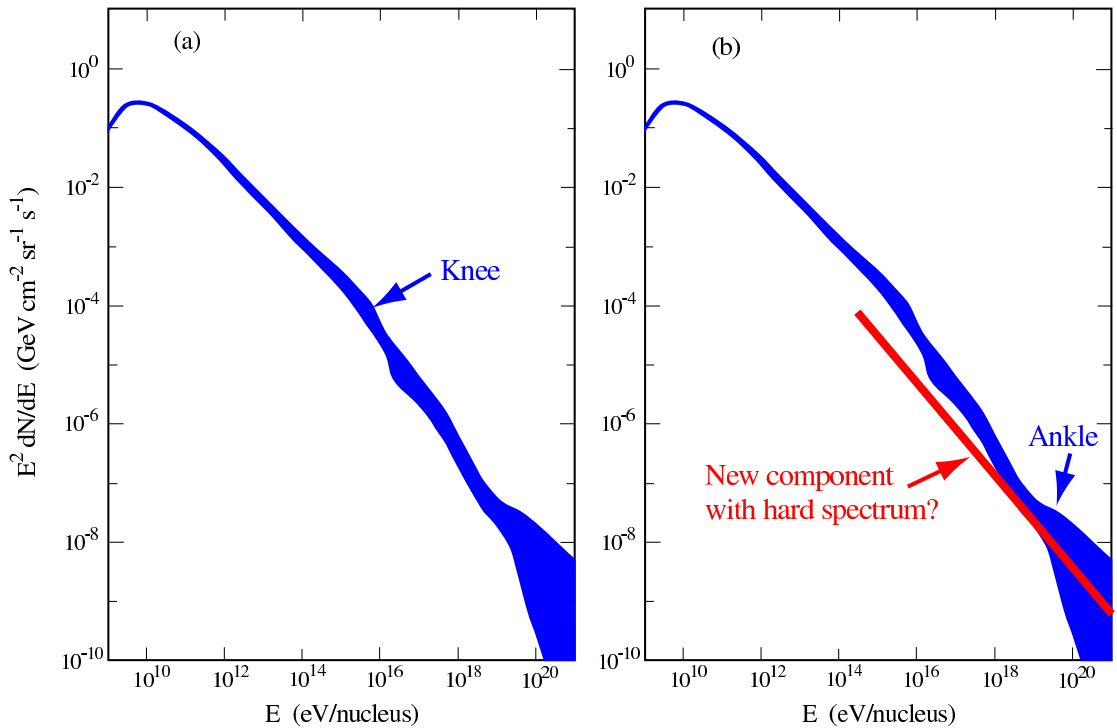}
\caption{At the energies of interest here, the cosmic ray spectrum consists of a sequence of 3 power laws. The first two are separated by the ``knee" (left panel), the second and third by the "ankle". There is evidence that the cosmic rays beyond the ankle are a new population of particles produced in extragalactic sources; see right panel.}
\label{knee-ankle}
\end{figure}

Models for the origin of the highest energy cosmic rays fall into two categories, top-down and bottom-up. In top-down models it is assumed that the cosmic rays are the decay products of cosmological remnants or topological defects associated, for instance, with Grand Unified theories with unification energy $M_{GUT} \sim 10^{24}\rm\,eV$. These models predict neutrino fluxes most likely within reach of first-generation telescopes such as AMANDA, and certainly detectable by future kilometer-scale neutrino observatories\cite{semikoz}. They have not been observed.

In bottom-up scenarios it is assumed that cosmic rays originate in cosmic accelerators. Accelerating particles to TeV energy and above requires massive bulk flows of relativistic charged particles. These are likely to originate from the exceptional gravitational forces  in the vicinity of black holes. Gravity powers large electric currents that create the opportunity for particle acceleration by shocks, a mechanism familiar from solar flares where particles are accelerated to $10$\,GeV. It is a fact that black holes accelerate electrons to high energy; astronomers observe them indirectly by their synchrotron radiation. We know that they accelerate protons because we detect them as cosmic rays. Because they are charged, protons are deflected by interstellar magnetic fields; cosmic rays do not reveal their sources. This is the cosmic ray puzzle.

Examples of candidate black holes include the dense cores of exploding stars, inflows onto supermassive black holes at the centers of active galaxies and annihilating black holes or neutron stars. Before leaving the source, accelerated particles pass through intense radiation fields or dense clouds of gas surrounding the black hole. This results in interactions producing pions decaying into secondary photons and neutrinos that accompany the primary cosmic ray beam as illustrated in Fig.\,2.
%
\begin{figure}[!h]
\centering\leavevmode
\includegraphics[width=4.25in]{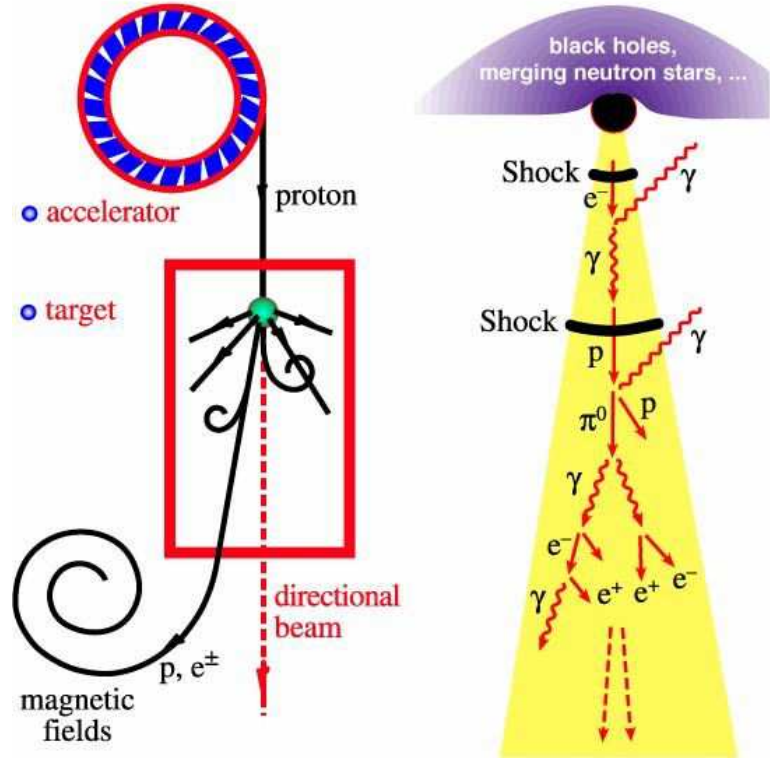}
\caption{Cosmic beam dump exits: sketch of cosmic ray accelerator producing photons. The charged pions that are inevitably produced along with the neutral pions will decay into neutrinos.}
\label{nubeams}
\end{figure}
 How many neutrinos are produced in association with the cosmic ray beam? 
 The answer to this question provides one rationale for building
 kilometer-scale  neutrino detectors~\cite{PR}. For orientation,
 consider a neutrino beam produced at an accelerator laboratory. 
 Here the target and the beam dump absorb all parent protons as
 well as the secondary electromagnetic and hadronic showers. 
 Only neutrinos exit the dump. If Nature constructed such a
 ``hidden source'' in the heavens, conventional astronomy would
 not reveal it.  Cosmic ray sources must be at least partially
 transparent to protons.  Sources transparent only to neutrinos
 may exist, but they cannot be cosmic-ray sources. 

 A generic ``transparent'' source can be imagined as follows: protons
 are accelerated in a region of high magnetic fields where they
 interact with photons and generate neutral and charged pions.
 The most important process is 
 $p + \gamma \rightarrow \Delta^+ \rightarrow \pi^0 + p$
 and
 $p + \gamma \rightarrow \Delta^+ \rightarrow \pi^+ + n$.
 While the secondary protons may remain trapped in the acceleration
 region, roughly equal numbers of neutrons and decay products of 
 neutral and charged pions escape. The energy escaping the source
 is therefore distributed among cosmic rays, gamma rays and
 neutrinos produced by the decay of neutrons, neutral pions and
 charged pions, respectively. 
 The neutrino flux from a generic transparent cosmic ray source is
 often referred to as the Waxman-Bahcall flux~\cite{wb1}. 
 It is easy to calculate and the derivation is revealing.

 Figure 1b shows a fit to the observed spectrum above the
 ``ankle" that can be used to derive the total energy in extragalactic
 cosmic rays. The flux above the ankle is often summarized as
 ``one $10^{19}$\,eV particle per kilometer square per year per
 steradian". This can be translated into an energy flux
\begin{eqnarray*}
E \left\{ E{dN\over dE} \right\} = {10^{19}\,{\rm eV} \over \rm (10^{10}\,cm^2)(3\times 10^7\,sec) \, sr}
= 3\times 10^{-8}\rm\, GeV\ cm^{-2} \, s^{-1} \, sr^{-1} \,.
\end{eqnarray*}
>From this we can derive the energy density $\rho_E$ in cosmic rays using the relation that flux${}={}$velocity${}\times{}$density, or
\[
4\pi \int  dE \left\{ E{dN\over dE} \right\} =  c\rho_E\,.
\]
We obtain
\[
\rho_E = {4\pi\over c} \int_{E_{\rm min}}^{E_{\rm max}} {3\times 10^{-8}\over E} dE \, {\rm {GeV\over cm^3}} \simeq 10^{-19} \, {\rm {TeV\over cm^3}} \,,
\]
taking the extreme energies of the accelerator(s) to be $E_{\rm max} / E_{\rm min} \simeq 10^3$.

The energy content derived ``professionally" by integrating the spectrum in Fig.~2b assuming an $E^{-2}$ 
energy spectrum, typical of shock acceleration, with a GZK cutoff  is $\sim 3 \times 10^{-19}\rm\,erg\ cm^{-3}$. 
This is within a factor of our back-of-the-envelope estimate (1\,TeV = 1.6\,erg). 
The power required for a population of sources to generate this energy density over 
the Hubble time of $10^{10}$\,years is $\sim 3 \times 10^{37}\rm\,erg\ s^{-1}$ per (Mpc)$^3$ or, 
as often quoted in the literature, $\sim 5\times10^{44}\rm\,TeV$ per (Mpc)$^3$ per year. 
This works out to\cite{TKG}
\begin{itemize}
\item $\sim 3 \times 10^{39}\rm\,erg\ s^{-1}$ per galaxy,
\item $\sim 3 \times 10^{42}\rm\,erg\ s^{-1}$ per cluster of galaxies,
\item $\sim 2 \times 10^{44}\rm\,erg\ s^{-1}$ per active galaxy, or
\item $\sim 2 \times 10^{52}$\,erg per cosmological gamma ray burst.
\end{itemize}
The coincidence between these numbers and the observed output in electromagnetic energy of these 
sources explains why they have emerged as the leading candidates for the cosmic ray accelerators. 
The coincidence is consistent with the relationship between cosmic rays and photons built into the ``transparent" 
source. In the photoproduction processes roughly equal energy goes into the secondary neutrons, neutral 
and charged pions whose energy ends up in cosmic rays, gamma rays and neutrinos, respectively.

We therefore conclude that the same energy density  of $\rho_E \sim 3 \times 10^{-19}\rm\,erg\
 cm^{-3}$, observed in cosmic rays and electromagnetic energy, ends up in neutrinos with a spectrum $E_\nu dN / dE_{\nu}  \sim E^{-\gamma}\rm\, cm^{-2}\, s^{-1}\, sr^{-1}$ that continues up to a maximum energy $E_{\rm max}$. The neutrino flux follows from the relation
%
$ \int E_\nu dN / dE_{\nu}  =  c \rho_E / 4\pi  $.
%
For $\gamma = 1$ and $E_{\rm max} = 10^8$\,GeV, the generic source of the highest energy cosmic rays produces a flux of $ {E_\nu}^2 dN / dE_{\nu}  \sim 5 \times 10^{-8}\rm\, GeV \,cm^{-2}\, s^{-1}\, sr^{-1} $.

There are several ways to sharpen  this qualitative prediction:
\begin{itemize} 
\item The derivation fails to take into account that there are more UHE
  cosmic rays in the Universe than observed at Earth because
  of the GZK-effect and it also neglects the evolution of the
  sources with redshift. This increases the neutrino flux, which
  we normalized to the observed spectrum only, by a factor
  $d_H/d_{\rm CMB}$, the ratio of the Hubble radius to the
  average attenuation length of the cosmic rays propagating in
  the cosmic microwave background.
\item For proton-$\gamma$ interactions muon neutrinos (and
  antineutrinos) receive only 1/2 of the energy of the charged pion in
  the decay chain $\pi^+\rightarrow \mu^+ +\nu_{\mu}\rightarrow e^+
  +\nu_e +\bar{\nu}_{\mu} +\nu_{\mu}$ assuming that the energy is
  equally shared between the 4 leptons. Furthermore half the muon
  neutrinos oscillate into tau neutrinos over cosmic distances. In further calculations we will focus on the muon flux here.
\end{itemize}
In summary,
\begin{equation}
E_\nu{dN_\nu\over dE_\nu} = {1\over2} \times {1\over2} 
\times E{dN_{\rm CR}\over dE} \times {d_H\over d_{\rm CMB}} 
\simeq E{dN_{\rm CR}\over dE}
\end{equation}
 In practice, the corrections approximately cancel. The precise value of the energy where the transition from galactic to
 extragalactic sources occurs represents another source of uncertainty that has been extensively debated~\cite{ringwald}. A transition at a lower energy significantly increases the energy in the extragalactic component and results in an enhancement of the associated neutrino flux.

Waxman and Bahcall referred to their flux as a bound in part because in
reality more energy is transferred to the neutron than to the charged
pion in the source, in the case of the photoproduction reaction $p +
\gamma \rightarrow \Delta^+ \rightarrow \pi^+ + n$ four times more.
Therefore
\begin{equation}
E_\nu{dN_\nu\over dE_\nu} = {1\over4} E{dN_{\rm CR}\over dE} \, .
\end{equation}
 In the end we estimate that the muon-neutrino flux associated with
 the sources of the highest energy cosmic rays is loosely confined
 to the range $ {E_\nu}^2 dN / dE_{\nu}= 1\sim 5 \times 10^{-8}\rm\,
 GeV \,cm^{-2}\, s^{-1}\, sr^{-1} $ depending on the cosmological
 evolution of the cosmic ray sources. 

The anticipated neutrino flux thus obtained has to be compared with the sensitivity 
of $8.9 \times 10^{-8}\rm\, GeV\ cm^{-2}\, s^{-1}\,sr^{-1}$ reached after the first 
4 years of operation of the completed AMANDA detector in 2000--2003~\cite{pune}. 
The analysis of the data has not been completed, but a limit of  
$2 \times 10^{-7}\rm\,GeV\ cm^{-2}\,s^{-1}\,sr^{-1}$ has been obtained with a single year 
of data\cite{HS}. On the other hand, after three years of operation IceCube will 
reach a diffuse flux limit of 
$E_{\nu}^2 dN / dE_{\nu} = 2\,{\sim}\, 7 \times 10^{-9}\rm\,GeV \,cm^{-2}\, s^{-1}\, sr^{-1}$. 
The exact value of the IceCube sensitivity depends on the magnitude of the dominant high energy 
neutrino background from the prompt decay of atmospheric charmed particles\cite{ice3}. The level 
of this background is difficult to anticipate theoretically and little accelerator data is available 
in the energy and Feynman-x range of interest\cite{ingelman}.

\begin{figure}[t]
\centering\leavevmode
\includegraphics[width=4.5in]{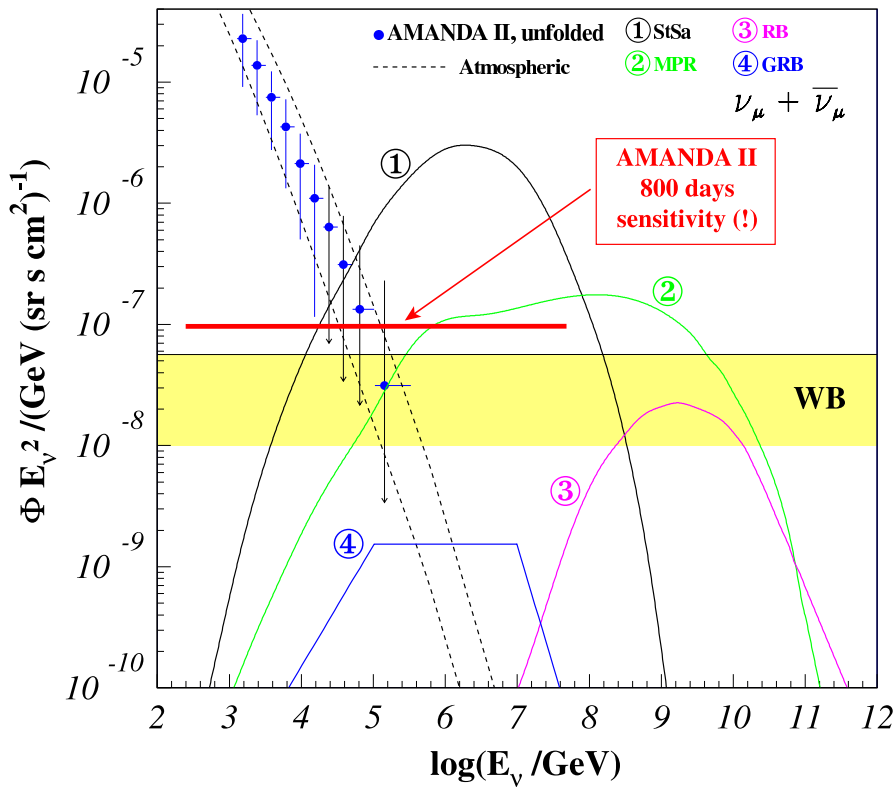}
\caption{Our estimate of the flux of neutrinos associated with the
 sources of the highest energy cosmic rays (the shaded range labeled WB)
 is compared to the sensitivity of the AMANDA experiment reached with
 800 days of data. Also shown are fluxes predicted by specific models
 of cosmic ray accelerators: active galaxies labeled StSa\protect\cite{agn}
 and MPR\protect\cite{MPR}, gamma ray bursts\protect\cite{guetta} and
 the diffuse flux produced by cosmic ray producing active galaxies on
 microwave photons\protect\cite{RB} labelled RB. Data for the background
 atmospheric neutrino flux are from the AMANDA experiment.}
\label{fig:diffuse_incl_osc_gs_ai_new.eps}
\end{figure}

The observed event rate is obtained by folding the cosmic flux predicted with the probability that 
the neutrino is actually detected in a high energy neutrino telescope; only one in a million neutrinos 
of TeV energy interact and produce a muon that reaches the detector. This probability is given by the 
ratio of the muon and neutrino interaction lengths in the detector medium, 
$\lambda_\mu / \lambda_\nu$\cite{PR} and therefore depends on energy. For the flux range estimated above we anticipate $20 \,{\sim}\, 100$ 
detected muon neutrinos per km$^2$ per year. Given that its effective area for muon neutrinos 
exceeds 1\,km$^2$ and that equal fluxes of electron and tau neutrinos are expected, a neutrino signal 
at the ``Waxman-Bahcall" level could result in the observation of several hundred high-energy neutrinos of extraterrestrial origin per year in IceCube~\cite{ice3}. 
Model calculations assuming that active galaxies or gamma-ray bursts are the actual sources of cosmic 
rays yield similar event rates than the generic energetics estimate presented.

Gamma ray bursts (GRB), outshining the entire Universe for the duration of the burst, are perhaps the best motivated sources of high-energy neutrinos\cite{waxmanbahcall, mostlum2, mostlum3}. The collapse of massive stars to a black hole has emerged as the likely origin of the "long" GRB with durations of tens of seconds. In the collapse a fireball is produced which expands with a highly relativistic velocity powered by radiation pressure. The fireball eventually runs into the stellar material that is still accreting onto the black hole. If it successfully punctures through this stellar envelope the fireball emerges to produce a GRB. While the energy transferred to highly relativistic electrons is thus observed in the form of radiation, it is a matter of speculation how much energy is transferred to protons.

The assumption that GRB are the sources of the highest energy cosmic rays does determine the energy of the fireball baryons. Accommodating the observed cosmic ray spectrum of extragalactic cosmic rays requires roughly equal efficiency for conversion of fireball energy into the kinetic energy of protons and electrons. In this scenario the production of $100 \sim 1000$\,TeV neutrinos in the GRB fireball is a robust prediction because neutrinos are inevitably produced in interactions of accelerated protons with fireball photons. Estimates of the flux\cite{guetta} point again at the necessity of a kilometer-cubed neutrino detector, in agreement with the generic energetics estimates previously presented. Studies of active galaxies as sources of cosmic rays lead to similar conclusions\cite{agn}.

The case for kilometer-scale detectors also emerges from consideration of  ``guaranteed'' cosmic fluxes. Neutrino fluxes are guaranteed when both the accelerator and the pion producing target material can be identified. We mention three examples. The extragalactic cosmic rays produce $ \sim$ 1 event per km$^2$\,year in interactions with cosmic microwave photons\cite{cos}.  Supernovae producing  cosmic rays in the dense star formation regions of starburst galaxies form a hidden source of neutrinos within reach of IceCube\cite{loeb&wax}.  Galactic cosmic rays interact with hydrogen in the disk to generate an observable neutrino flux in a kilometer-scale detector\cite{plane}.

\section{Cosmic Neutrinos Associated with Galactic Cosmic Rays}

 In the previous section we made an estimate of the neutrino flux
 from generic accelerators producing the highest energy cosmic rays.
 We can perform a similar analysis for the galactic cosmic rays by
 calculating the energy density corresponding to the flux shown in
 Fig.\,1a. The answer is that $\rho_E \sim 10^{-12}$\,erg\,cm$^{-3}$.
 This is also the value of the corresponding energy density $B^2/8\pi$
 of the microgauss magnetic field in the galaxy. The power needed to maintain
 this energy density is $10^{-26}$\,erg/cm$^3$s given that the
 average containment time of the cosmic rays in our galaxy is
 $3\times10^6$\,years. For a nominal volume of the galactic disk
 of $10^{67}$\,cm$^3$ this requires an accelerator delivering
 $10^{41}$\,erg/s. This happens to be 10\% of the power produced by supernovae releasing
 $10 ^{51}\,$erg every 30 years. The coincidence is the basis for
 the idea that shocks produced by supernovae exploding into the
 interstellar medium are the origin of the galactic cosmic
 rays.
  
With recent observations\cite{hess} of the supernova remnant
 RX J1713.7-3946 the H.E.S.S. array of atmospheric Cherenkov telescopes obtained circumstantial evidence for cosmic ray acceleration and, if confirmed, identified a guaranteed source of
 cosmic neutrinos\cite{alvarezhalzen}. With RX J1713.7-3946, H.E.S.S.  may have detected
 the first site where protons are accelerated to energies
 typical of the main component of the galactic cosmic rays\cite{ADV94}. Although
 the resolved image of the source (the first ever at TeV energies!)
 reveals TeV gamma ray emission from the whole supernova remnant,
 it shows a clear increase of the flux in the directions of known
 molecular clouds. This suggests
 the possibility that protons,
 shock accelerated in the supernova remnant, interact with the dense
 clouds to produce neutral pions that are the source of the observed
 increase of the TeV photon signal. The image shows filaments of
 high magnetic fields consistent with the requirements for
 acceleration to the energies observed. Furthermore, the high
 statistics data for the flux are power-law behaved over a large
 range of energies without any indication of a cutoff characteristic
 of synchrotron or inverse-Compton sources. Follow-up observations
 of the source in radio-waves and X-rays have failed to identify the
 population of electrons required to generate TeV photons by purely
 electromagnetic processes; for a detailed discussion see \cite{hiraga}. On the theoretical side, the large B-fields suppress the ratio of photons produced by the inverse Compton relative to the synchrotron. Fitting the data by purely electromagnetic processes is therefore challenging but, apparently, not impossible\cite{hiraga}.

Gamma ray telescopes have therefore not succeeded in finding the smoking gun for the supernova origin of the galactic cosmic rays. If the TeV flux of RX J1713.7-3946 is of neutral pion origin, then the accompanying charged pions will produce a guaranteed neutrino flux of roughly 10 muon-type neutrinos per kilometer-squared per year\cite{alvarezhalzen} and produce incontrovertible evidence for cosmic ray acceleration. From a variety of such sources we can therefore expect event rates of cosmic neutrinos of galactic origin similar to those estimated for extragalactic neutrinos in the previous section. Supernovae associated with molecular clouds are a common feature of associations of OB stars that exist throughout the galactic plane.

It is important to realize that there is a robust relation between the neutrino and gamma flux emitted by cosmic ray accelerators\cite{alvarezhalzen}. The $\nu_\mu + \bar\nu_\mu$ neutrino flux ($dN_\nu/dE_\nu$) produced by the decay of charged pions in the source can be derived from the observed gamma ray flux by energy conservation:
\begin{equation}
\int_{E_{\gamma}^{\rm min}}^{E_{\gamma}^{\rm max}}
E_\gamma {dN_\gamma\over dE_\gamma} dE_\gamma = K
\int_{E_{\nu}^{\rm min}}^{E_{\nu}^{\rm max}} E_\nu {dN_\nu\over dE_\nu} dE_\nu
\label{conservation}
\end{equation}
where ${E_{\gamma}^{\rm min}}$ ($E_{\gamma}^{\rm max}$) is the minimum (maximum) energy of the photons that have a hadronic origin. ${E_{\nu}^{\rm min}}$ and ${E_{\nu}^{\rm max}}$ are the corresponding minimum and maximum energy of the neutrinos. The factor $K$ depends on whether the $\pi^0$'s are of $pp$ or $p\gamma$ origin. Its value can be obtained from routine particle physics. In $pp$ interactions 1/3 of the proton energy goes into each pion flavor. In the pion-to-muon-to-electron decay chain 2 muon-neutrinos are produced with energy $E_\pi/4$ for every photon with energy $E_\pi/2$. Therefore the energy in neutrinos matches the energy in photons and $K=1$. The flux has to be reduced by a factor 2 because of oscillations. For $p\gamma$ interactions $K=1/4$. The estimate should be considered a lower limit because the observed photon flux to which the calculation is normalized may have been attenuated by absorption in the source or in the interstellar medium. 

In the case of supernova remnants the calculation of the neutrino flux can be performed on the back-of-the-envelope. Let's specialize to TeV photons.  From a source such as RX J1713.7-3946 a flux of $10^{-11}$ photons per $cm^2$ second is detected. This is consistent with theoretical expectations. As previously pointed out, a few supernovae per century transferring a fraction $\sim0.1$ of their energy, or about W$_{CR} = 10^{50}$\,erg, into the acceleration of cosmic rays can accommodate the observed flux up to the ``knee" in the spectrum. The acceleration takes place in the high magnetic fields created in the shock expanding into the interstellar medium. In the interaction of the shocked protons with the interstellar proton density of $n \sim 1$\,cm$^{-3}$, neutral pions are produced decaying into TeV photons close to the observed rate, or\cite{ADV94}
\begin{equation}
E{dN_{\gamma}\over dE} (>E) = 10^{-11} \: ({photons\over \rm cm^2 s}) ({W_{CR}\over \rm 10^{50} erg}) ({n\over \rm 1 cm^3}) ({d\over \rm 1 kpc})^{-2}
\label{galactic1}
\end{equation}
Here $d$ is roughly the distance appropriate for RX J1713.7-3946 and we therefore obtain a flux consistent with the H.E.S.S. observation. We can finesse the more formal derivation by simply assuming that each TeV gamma ray is accompanied by a neutrino from a charged pion to obtain an event rate of 3 detected neutrinos per decade of energy per km$^2$ year, a result readily obtained from the relation
\begin{equation}
{dN_{events}\over d(lnE)} (>E) = 10^{-11} \: ({photons\over \rm cm^2 \: s}) \: area \: time \: ({\lambda_\mu\over \rm \lambda_\nu}),
\label{galactic2}
\end{equation}
where the last factor represents, as before, the probability that the neutrino is detected. It is approximately $10^{-6}$ for the TeV energy considered here. From several such sources IceCube will detect a flux of neutrinos similar to the one associated with extragalactic sources.
 
In summary, the energetics of galactic as well as extragalactic cosmic rays points at the necessity to build kilometer-scale detectors to observe the associated neutrino fluxes that will reveal the sources. The case for doing neutrino astronomy with kilometer-scale instruments can also be made in other ways\cite{PR} and, as is usually the case, the estimates of the neutrino fluxes pointing at the necessity of such detectors are likely to be optimistic.

\section*{Acknowledgments}
I thank my IceCube collaborators as well as Concha Gonzalez-Garcia and Tom Gaisser for discussions. This research was supported in part by the National Science Foundation under Grant No.~OPP-0236449, in part by the U.S.~Department of Energy under Grant No.~DE-FG02-95ER40896, and in part by the University of Wisconsin Research Committee with funds granted by the Wisconsin Alumni Research Foundation.

\end{document}